\documentclass[aps,prd,a4paper,showpacs,tightenlines,showkeys,preprintnumbers,amsmath,amssymb,12pt]{revtex4}
\usepackage[latin1]{inputenc}
\usepackage{amsmath}
\usepackage{amsfonts}
\usepackage{amssymb}
\usepackage{amssymb}
\usepackage{epstopdf}
\usepackage{graphicx}
\usepackage{bm}
\usepackage{dcolumn}
\usepackage{color}
\usepackage{pst-plot}

\newcommand{\beq}{\begin{equation}}
\newcommand{\eeq}{\end{equation}}
\newcommand{\bea}{\begin{eqnarray}}
\newcommand{\eea}{\end{eqnarray}}

\begin{document}
\title{The final stage of  gravitationally collapsed  thick matter layers}

\author{ Piero Nicolini$^1$\email{nicolini@fias.uni-frankfurt.de}, Alessio Orlandi$^2$
\email{orlandi@bo.infn.it} and Euro Spallucci$^3$\email{spallucci@ts.infn.it}}
\affiliation{%$^1$Department of Physics, Loyola Marymount University, Los Angeles, CA\\
%\affiliation{
$^1$Frankfurt Institute for Advanced Studies (FIAS) and Institut 
f\"ur Theoretische Physik,
 Johann Wolfgang Goethe-Universit\"at, Frankfurt am Main, Germany\\
$^2$Dipartimento di Fisica, Universit\`a di Bologna and INFN, Sezione di Bologna, 
Italy\\
%}\affiliation{
$^3$Dipartimento di Fisica, Universit\`a di Trieste and INFN, Sezione di Trieste, Italy}

 %%%%%%%%%%%%%%%%%%%%%%%%%%%%%%%%%%%%%%%%%%%%%%%%%%%%%%%

\begin{abstract}
In the presence of a \textit{minimal length} physical objects cannot collapse
to an infinite density, singular, matter point. In this note we consider
the possible final stage of the gravitational collapse of ``thick'' matter
layers. The energy momentum tensor we choose to model these shell-like objects 
is a proper modification of the source for ``non-commutative geometry inspired'',
regular black holes. By using higher momenta of Gaussian distribution to localize
matter at finite distance from the origin,
we obtain new solutions of the Einstein's equation which smoothly
interpolates between Minkowski geometry near the center of the shell
and Schwarzschild spacetime far away from the matter layer. The metric is
curvature singularity free.
Black hole type solutions exist only for ``heavy'' shells, i.e.
 $M\ge M_{e}$, where $M_{e}$ is the mass of the
extremal configuration. 
We determine the Hawking temperature and a modified Area Law
taking into account the extended nature of the source.
\end{abstract}

%%%%%%%%%%%%%%%%%%%%%%%%%%%%%%%%%%%%%%%%%%%%%%%%%%


\maketitle

\section{Introduction}
Relativistic, self-gravitating matter shells have been throughly investigated in 
different sectors of theoretical physics: ``... from cosmic inflation to hadronic bags'' 
\cite{Aurilia:1987cp}.
Remarkable applications of gravitational shell models  can be found in the framework
of inflationary cosmology, where, both the ``birth'' and the evolution
of vacuum bubbles can be effectively described
in terms of the dynamics of the boundary surface engulfing a false vacuum domain
\cite{Aurilia:1987cp,Aurilia:1984cm,Blau:1986cw,Farhi:1989yr}. 
Matter shells  in
general relativity are modeled as ``zero-thickness'' membranes endowed with some 
characteristic tension determined by the underlying classical, or quantum, physics.
Neglecting the real width of the mass-energy distribution affects Einstein equations
by introducing a surface of discontinuity in the background spacetime. 
 This approximation  allows
to encode all the dynamics in the matching condition between the inner and outer
geometries \cite{Israel:1966rt}.
Furthermore, contracting matter shells provide useful analytic toy-models of collapsing
massive  bodies leading to black hole formation. In the spherically symmetric case the 
only dynamical degree of freedom is given by the shell radius and the system can be 
quantized according with the standard principle of quantum mechanics.
In this framework,  self-gravitating quantum  shells
open a window  over the still ``murky'' quantum features of evaporating 
mini black holes 
\cite{Ansoldi:1997hz,Alberghi:1999ig,Ansoldi:2002ju,Alberghi:2003ce,Alberghi:2006zy}.\\
In this paper we are going to investigate the static, final stage, 
of collapsed matter shell  in the presence of a fundamental
\textit{minimal length} forbidding the shell to contract into a singular matter point.  
The emergence of a minimal length, as a new fundamental constant of Nature on the same
ground as $c$ and $\hbar$, is a general feature of  different approaches to
quantum gravity  \cite{Garay:1994en,Calmet:2004mp,Fontanini:2005ik}.\\
In recent years we showed as the very concept of ``point-particle'' is 
meaningless if there exists a lower bound to physically measurable lengths. For instance, in a series of papers the repercussions of a natural ultraviolet cut off have been analyzed in the context of quantum field theory  \cite{Smailagic:2003rp,Smailagic:2003yb,Smailagic:2004yy,Spallucci:2006zj,Kober:2010um}. This basic notion 
can be also encoded into the Einstein equations through a proper choice of the energy
momentum tensor. 
The most remarkable outcomes of this procedure are the disappearance of curvature 
singularities
in the solutions of the Einstein equations \cite{Nicolini:2008aj}, a regular 
behavior of the
Hawking temperature which allows to determine the physical character
of the evaporation remnant 
\cite{Banerjee:2008gc,Casadio:2008qy,Myung:2006mz,Myung:2007qt,Myung:2007av}, and
a different form of the relation between entropy and area of the event horizon
reproducing the celebrated Area Law for large, semi-classical, black holes \cite{Brustein:2007jj,Nicolini:2010nb}.\\
The procedure has been applied in the whole array of physically meaningful black hole solutions like the neutral, non-rotating case \cite{Nicolini:2005zi,Nicolini:2005de,Nicolini:2005vd,Spallucci:2011rn} the charged, non-rotating case \cite{Ansoldi:2006vg,Spallucci:2008ez}, the ``dirty'', neutral, non-rotating case \cite{Nicolini:2009gw}, the spinning, neutral  \cite{kerrr} and charged case \cite{Modesto:2010rv}. In addition the regularity of the above metrics has been exploited in several complementary contexts like the decay of the deSitter universe by quantum black hole nucleation \cite{Mann:2011mm} and the case of dimensionally reduced spacetimes \cite{Mureika:2011py}. The procedure we followed has been also recognized as a special result \cite{Modesto:2010uh} of recently formulated non-local gravity proposals  \cite{Gaete:2010sp,Mureika:2010je} with important cross-fertilization in the AdS/CFT paradigm \cite{Nicolini:2011dp}.

The paper is organized as follows. In section \ref{fushell} we will derive  new 
solutions of Einstein equations describing, spherically symmetric, static,
 self-gravitating, thick matter
layers, in the presence of a fundamental minimal length. These type of objects
are modeled by means of a proper modification of the source for 
``non-commutative geometry inspired'', regular black holes. We replace
the Gaussian profile of mass-energy, peaked around the origin, with
higher moments of Gaussian distribution with maxima shifted at finite distance 
from the origin. The finite width of the distribution is determined by  the 
minimal length. In the case of lump-type objects, the minimal length measures 
the spread of mass-energy around the origin and removes the curvature 
singularity. 
For finite width matter layers, with energy density vanishing at short 
distance, not only the central curvature singularity is removed, but 
the extrinsic curvature discontinuity between ``inner'' and ``outer'' geometry
is cured, as well. Spacetime geometry is continuous and differentiable everywhere and
 smoothly interpolates between Minkowski metric near the center, 
and Schwarzschild spacetime far away from the matter layer. \\
As in the cases previously discussed, 
black hole type solutions exist only for ``heavy'' shells, i.e.
 $M\ge M_{e}$, where $M_{e}$ is the mass of the extremal configuration. ``Light''
 matter layers with $M < M_{e}$ will settle down in a smooth solitonic 
 type configuration with no horizons or curvature singularity.\\
In section \ref{thermo} we  study the thermodynamic 
properties of black hole solutions  and determine both the Hawking Temperature, $T_H$,
and the relation between entropy and area of the horizon. We recover the celebrated
Area Law in the limit of large, semi-classical, black holes. On a general ground,
we find the leading term is one fourth of the area but in units of an ``effective''
gravitational coupling constant, $G_N\left(\, r_+\,\right)$, depending from the
radius of black hole. For ``large'' $r_+$, the Newton Constant is recovered.
Finally, in section \ref{concl} we will draw the conclusions.

\section{Thick shells}
\label{fushell}

In a previous series of papers we solved the Einstein equations including the effects 
of a minimal length in a proper energy-momentum tensor
\begin{equation}
R_{\mu \nu} - \frac{1}{2}\, g_{\mu \nu}\,R = 8 \pi\,
G_N\, T_{\mu\nu},
\label{e1}
\end{equation}
where $ T_{\mu \nu}=\mathrm{diag} \left(\, \rho\ , p_r\ , p_\perp\ , p_\perp\,\right) $. 
The energy density is a minimal width Gaussian distribution
\begin{equation}
\rho=\rho_0\left(\, r\,\right)\equiv\frac{M}{(4\pi\theta)^{3/2}}\, e^{-r^2/4\theta},
\label{gaussprofile}
\end{equation}
representing a ``blob-like'' object, centred around the origin,
with a  characteristic extension given by $l\propto \sqrt{\theta}$. A simple way to derive the above energy profile is based on the following considerations. At classical level point-like objects in spherical coordinates are described by a profile
\begin{equation}
\rho_{\mathrm{cl}}\left(\, r\,\right)=\frac{M}{4\pi r^2}\, \delta\left(\, r\,\right)
\label{rosing}
\end{equation}
where $\delta\left(\, r\,\right)$ is the Dirac delta. We recall that a Dirac delta function can be represented as the derivative of a  Heaviside step-function $\Theta(r)$
\begin{equation}
 \delta\left(\, r\,\right)=\frac{d}{dr}\Theta\left(\, r\,\right).
\end{equation}
However in the presence of a minimal length the very concept of sharp step is no longer meaningful. Rather we expect the local loss of resolution to sweeten the step. Accordingly it can be shown that in the framework of a minimal length a modified step function can be defined through an integral representation of the Heaviside function without taking the limit $\sqrt{\theta}\to 0$ \cite{Mureika:2011hg}
 \begin{eqnarray}
\Theta(r)\rightarrow P_0(r)&=&\frac{1}{\sqrt{4\pi} \theta^{3/2}} \int_0 ^r x^2 e^{-x^2/4\theta}dx	 \label{newstep} %\\    &=&\frac{\gamma(3/2; r^2/4\theta)}{\Gamma(3/2)}. \nonumber
\end{eqnarray}
%
%where $\gamma(a; x)\equiv\int_0^x t^{a-1}e^{-t}$ is the incomplete Gamma function.
The Gaussian profile (\ref{gaussprofile}) is therefore obtained as 
\begin{equation}
\rho_{0}\left(\, r\,\right)=\frac{M}{4\pi r^2}\, \frac{d}{dr} P_0\left(\, r\,\right)
\label{rhozero}
\end{equation}
In order solve Einstein equations, we need to determine the remaining components of the energy-momentum tensor.  The radial pressure $p_r$ is fixed by the equation of state 
$p_r=-\rho $ reproducing the deSitter ``vacuum'' equation of state at short distance. 
This is a key feature to build up a regular, stable configuration, where the
negative pressure balances the gravitational pull. In other words, the singularity
theorem is evaded by a violation of null energy condition triggered by the short
distance vacuum fluctuations.\\  
Finally, the tangential pressure $p_\perp $ is obtained in terms of $\rho $ by the divergence free condition $\nabla_\mu \, T^{\mu\nu}=0 $.\\
As the source is static and spherically symmetric, the line element can be cast in the 
form
\begin{equation}
ds^2=- f(r)\, dt^2 + \frac{dr^2}{f(r)}+r^2\Omega^2,
\label{e}
\end{equation}
with 
\begin{equation}
f(r)=1-\frac{2G_N\,  m(r)}{r}.
\label{f}
\end{equation}
The cumulative mass distribution $m(r)$ is given by
\begin{equation}
m(r) =  4\pi \int_0 ^r dr^\prime (r^\prime)^2\ \rho_0(r^\prime).
\label{mass}
\end{equation}
We notice that the parameter $M$ corresponds to the total mass energy of the system, 
namely
\begin{equation}
M = \lim_{r\to\infty}m(r).
\label{totalmass}
\end{equation}
The above profile cures the usual Dirac delta (singular) distribution associated to a
point-particle, and leads to a family of regular black hole  solutions 
\cite{Nicolini:2005zi,Nicolini:2005de,Nicolini:2005vd,Spallucci:2011rn,Ansoldi:2006vg,Spallucci:2008ez,
Nicolini:2009gw,kerrr,Modesto:2010rv,Mann:2011mm,Mureika:2011py}.\\  

\begin{figure}[htb]
\begin{center}
\includegraphics[height=7.5cm]{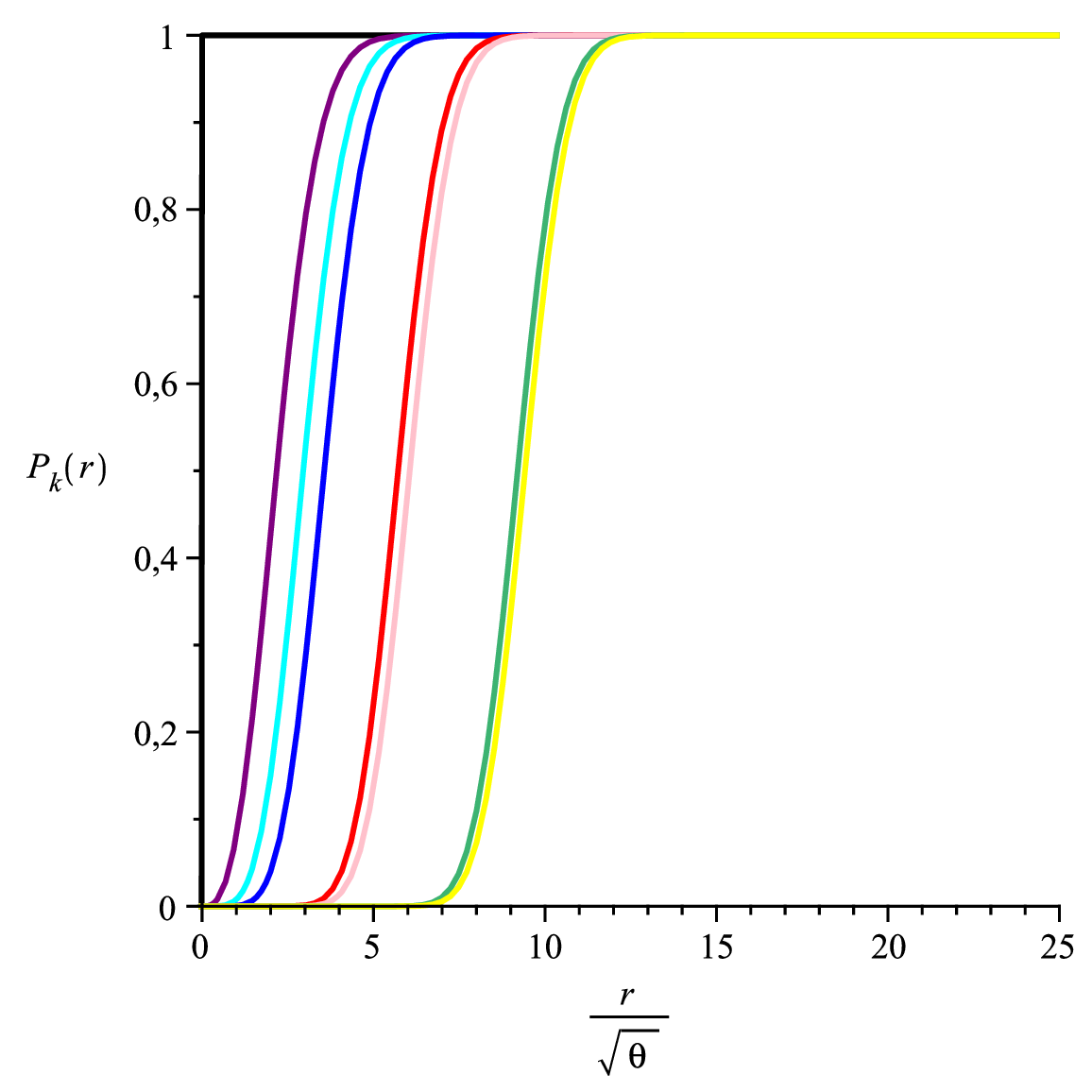}
\caption{\label{pkappa}
\emph{Plot of the function $P_k(r)$ as a function of $r/\sqrt{\theta}$ for
$k=0$ (purple)
$k=1$ (cyan),
$k=2$ (blue),
$k=7$ (red),
$k=8$ (pink),
$k=20$ (green),
$k=21$ (yellow).
For comparison, the Heaviside function is plotted in black.}}
\end{center}
\end{figure}

A crucial point at the basis of the above line of reasoning is the modification of the step function in the presence of a minimal length. We notice that the profile (\ref{newstep}) is just one of the possible choices one can have to account for the loss of resolution of the edge of the step \cite{Mureika:2011hg}. In other words there exist alternative representations of the Heaviside function for which one can deliberately  avoid the limit $\sqrt{\theta}\to 0$. For instance the family of distributions
 \begin{eqnarray}
\Theta(r)\rightarrow P_k(r)&=&\frac{1}{\sqrt{\pi} \theta^{k+3/2}}\frac{(k+1)!}{[2(k+1)]!} \int_0 ^r x^{2k+2} e^{-x^2/4\theta}dx	  %\\    &=&\frac{\gamma(3/2; r^2/4\theta)}{\Gamma(3/2)}. \nonumber
\end{eqnarray}
account for minimal length effects growing with the index $k$, where $k=0\ ,1\ , 2\, \dots$ is a natural number (see Fig. \ref{pkappa}). This can be seen by the limit 
$r/\sqrt{\theta}\gg 1$ 
\begin{equation}
P_k(r)\approx  1-\frac{4^{k+1}}{\sqrt{\pi}}\frac{(k+1)! }{[2(k+1)]!}  \left(\frac{r}{2\sqrt{\theta}}\right)^{2k+1} e^{-r^2/4\theta}	
\end{equation}
which shows that the distributions $P_k(r)$ reach the \textit{plateau} (\textit{i.e.} $P_k\approx 1$) more slowly as $k$ increases.

Here we want to explore the nature of the solutions emerging for the above admissible profiles for a modified step function.
%We are going now to adapt the above approach to an object which is \textit{shell-like}, rather than point-like. For this reason, we recall that the Gaussian distribution admits moments of all orders. The regular black holes we found from  Gaussian  distribution are just gravitational objects corresponding to the zero-th moment of the distribution. Here we want to explore the nature of the solutions emerging from higher moments of the Gaussian, namely normalized matter distributions of the kind
By replacing $P_0$ with a generic $P_k$ in (\ref{rhozero}), we derive the following energy density profile
\begin{equation}
 \rho_k\left(\, r\,\right)\equiv 
 M\,\frac{ r^{2k} e^{-r^2/4\theta}}{4^{k+2}\pi \theta^{k+3/2}\, 
 \Gamma\left(\, k+3/2\,\right)}
 \label{uno}
\end{equation}
which corresponds to higher moments of the Gaussian.
%where $k=0\ ,1\ , 2\, \dots$ is a natural number. 
%The regular black holes we found from  Gaussian  distribution are just gravitational objects corresponding to the zero-th moment of the distribution. 
%We are considering  the case of even moments only. This choice is motivated by the request of having a rapidly vanishing energy profile at the origin, corresponding to the case of ``empty'' shells. On the contrary odd moments may admit a linearly vanishing energy profile, which gives rise to a gravitational object which does not properly describe a matter shell. We stress that the choice of even moments does not modify any of the following conclusions for both the geometric and the thermodynamic properties of the solutions.  

\begin{figure}[htb]
\begin{center}
\includegraphics[height=5cm]{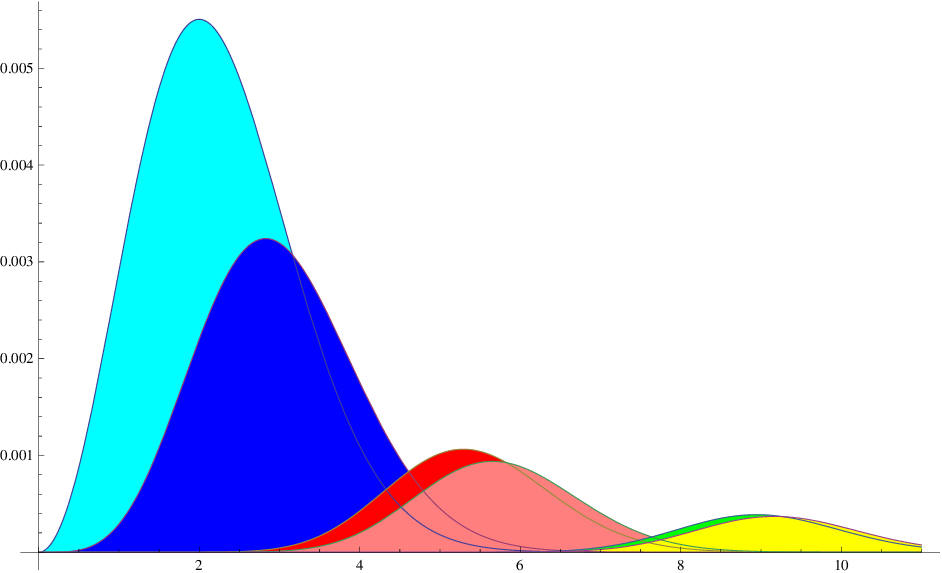}
\caption{\label{rhok}
\emph{Plot of the radial density profile $\rho(r)$ as a function of $r$ in $\sqrt{\theta}$-units for
$k=1$ (cyan),
$k=2$ (blue),
$k=7$ (red),
$k=8$ (pink),
$k=20$ (green),
$k=21$ (yellow).
The function $\rho(r)$ has been normalized so that its integration over a spherical 
volume gives 1.}}
\end{center}
\end{figure}

For $k=0$ the function $\rho_k$ turns into the Gaussian distribution, centred around 
the origin, 
while for $k\ge 1$ the matter distribution is more and more diluted near the origin, 
being peaked at $r_M= 2\sqrt{k\theta}$.  As a result the density function (\ref{uno}) 
describes a whole family of ``mass-degenerate'' shells, with the same  $M$, 
but concentrated at a distance given by $r_M$ (see Fig. \ref{rhok}). 
Eq. (\ref{uno}) discloses further properties of the minimal length, 
which assumes a new intriguing meaning. In the case of point-particle, $\sqrt{\theta}$ 
represents the spread of the object around the origin. For  matter shells, 
we see $\sqrt{\theta}$ relates to  the shell thickness or, in other words, a measure 
of the intrinsic fuzziness of the layer.  Just as a particle cannot be exactly localized 
at a single point, we cannot have zero-width layers, as well.  
Thus, as the matter distribution is smooth everywhere, there is no
discontinuity in the extrinsic curvature between the ``inner'' and outer ``geometry''.
No matching condition is required and we can look for a single, smooth, metric
inside, across and outside the matter layer.\\
We can define the  distance between two  shells corresponding to different moments
as the distance between the peaks:
\begin{equation}
 \frac{\Delta r_M}{2\sqrt\theta} = \sqrt{k+1} -\sqrt{k}
 \ .
\label{distance}
\end{equation}
We see that for higher moments  $\Delta r_M$  vanishes as we are at length scale
much larger than $\sqrt\theta$ and the relative distance cannot be resolved
anymore.\\
A stable solution of the Einstein equations can be obtained by sourcing the 
gravitational field by the energy momentum tensor of an anisotropic fluid:
%
%\begin{eqnarray}
% &&{\cal S}^0{}_0 = -\rho_k\\
%&&{\cal S}^r{}_r = p_r=- \rho_k\\
%&&{\cal S}^\theta{}_\theta = {\cal S}^\phi{}_\phi= p_t
%\end{eqnarray}
%
%where $p_r$, $p_t$ are the radial and tangential pressure, respectively. 
the choice $p_r=- \rho_k$ for the matter equation of state  allows to have
$g_{00}=-g_{rr}^{-1}$. Furthermore,  the hydrodynamic
equilibrium equation will give $p_\perp$ in terms of $\rho_k$, while the metric itself 
results to be independent
from $p_\perp$. The solution of the Einstein equations  reads in geometric units, 
$c=1$, $G_N=1$:
\begin{eqnarray}
&& ds^2=-\left(\, 1 -\frac{2m\left(\, r\,\right)}{r}\,\right)dt^2 
+\left(\, 1 -\frac{2m\left(\, r\,\right)}{r}\,\right)^{-1} dr^2 
+ r^2 d\Omega^2\label{due}\\
&& m\left(\, r\,\right)\equiv 4\pi
\int_0^r dr^\prime \, r^{\prime\, 2}\, \rho\left(\, r^\prime\,\right)
= M\,\frac{\gamma\left(\, k+3/2\ ; r^2/4\theta\,\right)}{ \Gamma\left(\, k+3/2\,\right)}
\label{tre}
\end{eqnarray}
where $\gamma\left(\, k+3/2\ ; r^2/4\theta\,\right)$ is the lower, incomplete, 
Euler Gamma Function.
\begin{figure}[htb]
\begin{center}
\includegraphics[height=5cm]{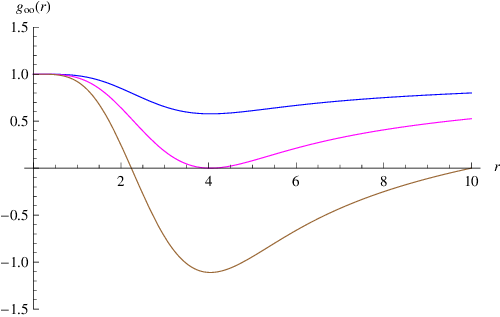}
  \caption{\label{g00}\emph{Plot of $g_{00}(r)$ as a function of $r$ in $\sqrt{\theta}$-units for fixed value of $k=1$ and different values of the mass $M$, i.e. $M=1$ (blue), $M=M_e$ (magenta) and $M=5$ (brown). Here $M_e \simeq 2.36976$ is the extremal mass for the case $k=1$.}}
\end{center}
\end{figure}
%
%\section{Regularity in the origin}
By taking into account the asymptotic form of 
$\gamma\left(\, k+3/2\ ; r^2/4\theta\,\right)$
for small argument, one finds that the metric (\ref{due}) is essentially flat near 
the origin, the contrary to what happens for the case $k=0$ in which a regular 
deSitter core forms. Indeed one finds
\begin{eqnarray}
&& f\left(\, r\,\right)\simeq 1 -\frac{1}{2k+3}\left[\frac{4M}{r\, \Gamma\left(\, k+3/2\,\right)}\right] 
\,\left(\, \frac{r^2}{4\theta}\,\right) ^{k+3/2}\ ,\\
&& ds^2=-\left(\, 1 -O\left(\, r^{2k+2}\,\right)   \,\right)\, dt^2 
+\left(\, 1 - O\left(\, r^{2k+2}\,\right)\,\right)^{-1}\, dr^2 
+ r^2 d\Omega^2\label{mink}
\ .
\end{eqnarray}
The asymptotic behavior (\ref{mink}) can be seen as a
generalization of the ``Gauss Theorem'' : inside an empty, classical, thin shell 
of matter the  Newtonian gravitational field is zero, i.e. spacetime
is flat. Introducing a finite width density,  smoothly decreasing both toward the 
origin and space-like infinity, causes small deviations from perfect flatness.\\
Going back to (\ref{due}),
the presence of event horizons is read from the zeroes of $g_{rr}^{-1}$ 
(see Fig. \ref{g00}).
 Due to the non-trivial
structure of the metric, it is more convenient to look at the ``horizon equation'' in 
the form
\begin{equation}
 M\equiv U\left(\, r_H\,\right), \ U\left(\, r_H\,\right)\equiv \frac{r_H}{2}
 \frac{\Gamma\left(\, k+3/2\,\right)}{ \gamma\left(\, k+3/2\ ; r^2_H/4\theta\,\right)}
\label{heq}
\ .
\end{equation}
The problem of finding the horizons in the metric (\ref{due}) is mapped in the
equivalent problem of determining the turning-points for the motion of a 
``test-particle'' of energy $M$ subject to the ``potential'' $ U\left(\, r_H\,\right)$.
\begin{figure}[htb]
\begin{center}
\includegraphics[height=5cm]{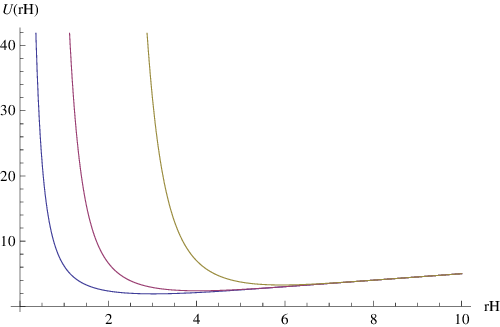}
\caption{\label{urho}\emph{Plot of $U(r_H)$ as a function of $r_H$ in $\sqrt{\theta}$-units for $k=0$ (blue), $k=1$ (magenta) and $k=4$ (brown).}}
\end{center}
\end{figure}
The intersection(s) between a line $M=\mathrm{const.}$ and the plot
of the function $U\left(\, r_H\,\right) $, in the plane $M$-$r_H$, represent the 
allowed radii of inner/outer event horizons (see Fig. \ref{urho}).  The asymptotic 
behavior of $U\left(\, r_H\,\right) $
is obtained from the corresponding approximate forms of $\gamma$:
\begin{eqnarray}
&& U\left(\, r_H\,\right)\approx \frac{r_H}{2}\ , \quad r_H>> \sqrt{\theta}\\
&& U\left(\, r_H\,\right)\approx\sqrt{\theta}\ \Gamma(k+5/2)\ \left(\frac{r^2_H}{4\theta}\right)^{-(k+1)}\ , \quad r_H<< \sqrt{\theta}
\ .
\end{eqnarray}
Thus, $U\left(\, r_H\,\right)$ is a convex function with a single minimum determined by 
the condition
\begin{equation}
\frac{dU}{dr_H}=0\longrightarrow \frac{r_e^{2k+3}}{2^{2k+2}\theta^{k+3/2}}=
\gamma\left(\, k+3/2\ ; r^2_e/4\theta\,\right)e^{r_e^2/4\theta}\label{extremal}
\ .
\end{equation}
The radius $r_e$ represents the size of an \textit{extremal} configuration with a 
couple of degenerate
 horizons: $r_-=r_+\equiv r_e$. Once $r_e$ is numerically determined from (\ref{extremal}), 
 one gets the corresponding mass $M_e$ from the potential:
\begin{equation}
 M_e=U\left(\, r_e\,\right)=U_{min.}
 \label{mex}
 \end{equation}
An order of magnitude estimate for  $r_e$ and  $M_e$ can be obtained by keeping
only the leading theta-term:
\begin{equation}
 r_e\propto \sqrt{\theta}\longrightarrow
     M_e\propto \frac{\sqrt\theta}{L_{Pl.}} \, M_{Pl.} 
\end{equation}
where $L_{Pl.}$ is the Planck length, i.e.,  $L_{Pl.}=M_{Pl.}^{-1}=\sqrt{G_N}$.
This  estimate suggests that (near/)extremal configurations are close to a full
 quantum gravity regime. As such, they are appropriate candidates to describe 
the end-point of the Hawking evaporation process where the semi-classical 
description breaks down.\\
In summary:
\begin{itemize}
\item
for $M>M_e$ we find a geometry  with non-coincident  inner and outer horizons  $r_-<r_+$. 
It is worth
to remark that there is no curvature singularity in $r=0$. 
Spacetime is flat near the origin.
\item
For $M=M_e$ we have an \textit{extremal} configuration with a pair of degenerate 
horizons, $r_-=r_+\equiv r_e$.
\item
For $M<M_e$ there are neither horizons nor curvature singularities. The metric is 
smooth and regular
everywhere.  The shell is too light and diluted to produce any relevant alteration 
of the spacetime fabric.  After collapse it will settle into a sort of solitonic
object with no horizon of curvature singularity.
\end{itemize}
Massive layers will produce horizons, but no curvature singularities. This is a crucial point which marks a departure with respect to all the existing literature. Our approach must not be confused with previous contributions in which generalized matter shells containing polytropic and Chaplygin gas cannot ultimately resolve the emergence of singularities \cite{Oh:2009cz}.

We can estimate the size of these objects by solving iteratively the equation
(\ref{heq}) and truncating the procedure at the first order in the expansion
parameter $\exp\left(\, -M^2/\theta\,\right)$:
\begin{equation}
 r_+\simeq 2M\left[1-\frac{(M^2/\theta)^{k+1/2}}{\Gamma(k+3/2)}\ e^{-M^2/\theta}\right].
 \end{equation}
This quantity can be compared with the shell \textit{mean radius}
\begin{equation}
\langle\, r\,\rangle \equiv \frac{4\pi}{M}\int_0^\infty dr r^2\, r\, \rho\left(\, r \,\right)=
2\sqrt{\theta}\ \frac{\Gamma(k+2)}{\Gamma(k+3/2)}
\ .
\end{equation}

We notice that for $k>>1$ the leading contribution to the ratio 
$r_+/ \langle\, r\,\rangle$ is given by
\begin{equation}
\frac{r_+}{ \langle\, r\,\rangle } \approx \frac{M}{\sqrt\theta}\ \frac{1}{\sqrt{k+1}}
\ .
\end{equation}
Thus, for an assigned $M$ the ratio decreases with $k$. The horizon radius
is determined by the total mass energy $M$ and is weakly $k$-dependent.
On the contrary, $ \langle\, r\,\rangle \propto \sqrt{k+1}\, \sqrt\theta $ and grows
with $k$, which means that for higher moments the horizon is surrounded by a cloud
of matter.

\section{Thermodynamics}
\label{thermo}
\begin{figure}[htb]
\begin{center}

\includegraphics[height=5cm]{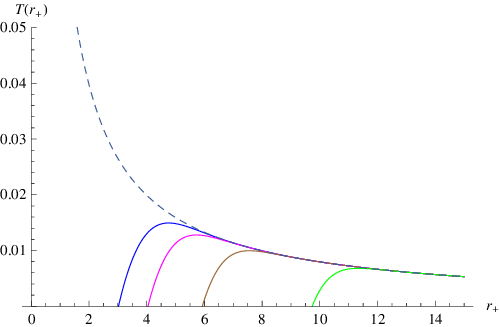}
  \caption{\emph{Plot of $T(r_+)$ as a function of $r_+$ in $\sqrt{\theta}$-units for $k=0$ (blue), $k=1$ (magenta) and $k=4$ (brown), $k=15$ (green). The dashed line represent the classical case, i.e. $\theta=0$.}}
  \label{shelltemp}
  
\end{center}
\end{figure}

Massive shells will collapse  into black holes described by the line element
(\ref{e}),(\ref{f}). This is not the end of story as these objects are
semi-clasically unstable under Hawking emission.
We are now ready to study the thermodynamic properties of these solutions starting. 
from the Hawking temperature:
\begin{equation}
 T_H=\frac{1}{4\pi r_+}\left[\, 1 
 -\frac{r_+^{2k+3}}{2^{2k+2}\theta^{k+3/2}}
 \frac{e^{-r_+^2/4\theta} }{ \gamma\left(\, k+3/2\ ; r_+^2/4\theta\,\right)}
 \,\right]
\label{th}
\ .
\end{equation}
 It can be easily  verified that $T_H$ is vanishing for the extremal configuration
 $ T_H\left(\, r_+=r_e\,\right)=0 $.\\
 For $k=0$ we obtain the temperature of the black hole as in 
\cite{Nicolini:2005vd}. The behavior of the temperature can be found 
in Fig. \ref{shelltemp}. 
We notice that the regularity of the manifold for any $k$ leads  to a cooling 
down of the horizon in the terminal phase of the Hawking process. At $r_+=r_{max}$ the presence of
a maximum temperature corresponds to an infinite discontinuity 
 the heat capacity which is usually interpreted as the signal of a ``change of state''
 for the system. For $r_+> r_{max}$ the black hole is thermodynamically unstable
 and increases its  temperature by radiating away its own mass.  After crossing
 $r_{max}$, i.e., for $r_+<r_{max}$ the black hole enters a stability phase asymptotically ending into
 a degenerate ( zero-temperature ) extremal configuration.
 By increasing $k$ we just lower down the maximum temperature. As a result the 
 solution is unaffected by any relevant quantum back reaction as already proved 
 for the case $k=0$ in \cite{Nicolini:2005vd}.
 \\
 A further important consequence of $T_H\le T_{max}$ is that ( in extra-dimensional
 models ) the black hole mainly radiates on the brane \cite{Nicolini:2011nz}, it never becomes
  hot enough to warm-up the bulk in a significant way.

\begin{table}[ht]

\centering
\caption{\label{Tab1}Some values of parameters of the matter shells}
    \begin{tabular}{||c | c | c | c | c || }
    \hline
     \hline
     & $r_e$ & $M_e$ & $T_{max}$ & $r_{max}$ \\
\hline
 \hline
    $k=0$ & 3.0224 & 1.9041 & 0.014937 & 4.76421 \\
\hline
    $k=1$ & 4.0431 & 2.3698 & 0.012783 & 5.72632 \\
\hline
    $k=4$ & 5.9269 & 3.2647 & 0.009960 & 7.56069 \\
\hline
    $k=15$ & 9.7347 & 5.1245 & 0.006799 & 11.3419 \\
\hline
 \hline
    \end{tabular}
    
    \end{table}

 The area-entropy law can be recovered from the relation
\begin{equation}
dM= T_H\, dS 
\label{primalegge}
\end{equation}
where $dS$ is the horizon entropy variation triggered by a variation $dM$ in the total 
mass energy $M$. In order to translate the first law of black hole thermodynamics 
(\ref{primalegge})
into a relation involving the area of the event horizon, we need to write mass 
energy variation as
 \begin{equation}
dM= \frac{\partial U}{\partial r_+}\, dr_+
\label{du}
\end{equation}
and to take into account that the minimum of $U\left(\, r_+\,\right)$ is the
mass of the extremal black hole.
Thus, when integrating (\ref{du}) the lower integration limit is the radius
of the extremal configuration
\begin{equation}
 S= \int^{r_+}_{r_e} dx\, \frac{1}{T_H}\frac{\partial U}{\partial x}=
 2\pi\, \Gamma\left(\, k+3/2\,\right)\, \int^{r_+}_{r_e} dr\, 
 \frac{r}{ \gamma\left(\, k+3/2\ ; r^2/4\theta\,\right)},
\label{entropy}
\end{equation}
where we have inserted  (\ref{th}) and (\ref{du}) into (\ref{entropy}). By performing the integral one gets
\begin{eqnarray}
S = && \pi\,\Gamma\left(\, k+3/2\,\right) \,
\left(\, \frac{r_+^2}{\gamma\left(\, k+3/2\ ; r_+^2/4\theta\,\right) }     
-\frac{r_e^2}{\gamma\left(\, k+3/2\ ; r_e^2/4\theta\,\right) } 
\,\right) + 
\nonumber\\
  && \pi \,\Gamma\left(\, k+3/2\,\right)\int^{r_+}_{r_e} dr \, r^2 
\frac{\gamma^\prime}{\gamma^2}.
\label{sh}
\end{eqnarray}
 The first term can be written in terms of the area of the event horizon as
 \begin{equation}
 S=\frac{1}{4G_N}\,\Gamma\left(\, k+3/2\,\right) \,
\left(\, \frac{A_H}{\gamma\left(\, k+3/2\ ; r_+^2/4\theta\,\right) }     
-\frac{A_e}{\gamma\left(\, k+3/2\ ; r_e^2/4\theta\,\right) } 
\,\right) + \dots \label{s2}
 \end{equation}
 and represents the ``Area Law'' in our case. We have re-inserted the Newton
 constant into (\ref{s2}) for reasons to become clear in a while. 
 The standard form, which is one fourth
 of the area, is recovered in the large black hole limit, $r_H>> \sqrt\theta$.\\
 Once (\ref{s2}) is written in natural units, we can define
 an \textit{effective} Newton constant as
 \begin{equation}
 G_N\rightarrow G_N\left(\, r_+\, \right)\equiv G_N\,
 \frac{\gamma\left(\, k+3/2\ ; r_+^2/4\theta\,\right)}{\Gamma\left(\, k+3/2\,\right)}
 \end{equation}
 and introduce a ``modified'' Area Law as
 \begin{equation}
 S\left(\, r_H\,\right)= \frac{\pi\,r^2_H}{4G_N\left(\, r_H\, \right)}. 
 \end{equation}   
 The interesting feature of $ G_N\left(\, r_H\, \right)$ is to be \textit{``asymptotically
 free''} in the sense that it is smaller than $G_N$, which represents the asymptotic
 value of the gravitational coupling for large black holes only
 \footnote{It is interesting to remark that if we apply the same reasoning to
 the Newton constant in (\ref{f}), and replace $r_H$ with the radial coordinate
 $r$, we find a radial distance dependent gravitational coupling, which is quickly
 vanishing in the limit $r\to 0$. This asymptotically-free behavior of our
 model, at short distance, provides an alternative explanation for the absence
 of curvature singularity in $r=0$.}.\\

 Finally, the second term in (\ref{sh}), gives exponentially small corrections
 to the leading area term. It can be analytically computed for $k>>1$, but it does not 
 introduce any relevant new effect.
\section{Conclusions}
\label{concl}

In this paper we derived new static, spherically symmetric regular solutions of Einstein equations, describing the final state of collapsing matter shells in the presence of an effective minimal length. This derivation is the result of a long path started with the quest of quantum gravity regularized black hole solutions \cite{Nicolini:2005zi,Nicolini:2005de,Nicolini:2005vd,Spallucci:2011rn,Ansoldi:2006vg,Spallucci:2008ez,
Nicolini:2009gw,kerrr,Modesto:2010rv,Mann:2011mm,Mureika:2011py,Modesto:2010uh,Nicolini:2008aj}. We showed that our previous discoveries of noncommutative geometry inspired solutions correspond to the simplest case within a larger class of regular gravitational objects. This class, the family of gravitational shells here presented, has peculiar properties which descend from the common key feature of the absence of curvature singularities. Such properties are: the existence of extremal configurations, characterized by a minimal mass below which horizon cannot form, even in the
     case of neutral, non-rotating black holes;  the occurrence of a phase transition from a thermodynamically unstable classical phase to a thermodynamically stable cooling down in the final stage of the horizon evaporation. As for the case of regular black holes, these matter shells correspond to final configurations of dynamical processes of gravitational collapse in which matter cannot be compressed below a fundamental length scale.
The above properties are also common to other models of quantum geometry, that have been derived by means of different formalisms. Specifically the shell profile of the energy density resembles what one has in the case of loop quantum black holes, with consequent thermodynamic similarities \cite{leo1,leo2,leo3,leo4}.

The study of these objects is far from being complete. We believe that these new solutions can have repercussions in a variety of fields. Here we just mention that these regular shells could affect the stability of the deSitter space, at least during inflationary epochs \cite{Mann:2011mm} as well as they could lead  to new insights in the physics of nuclear matter via the gauge/gravity duality paradigm \cite{Nicolini:2011dp}.
\begin{acknowledgments}
This work has been supported by the project ``Evaporation of microscopic black holes'' under the grant NI 1282/2-1 of the German Research Foundation (DFG), by the Helmholtz International Center for FAIR within the framework of the LOEWE program (Landesoffensive zur Entwicklung Wissenschaftlich-\"{O}konomischer Exzellenz) launched by the State of Hesse and in part by the European Cooperation in Science and Technology (COST) action MP0905 ``Black Holes in a Violent Universe''. A.O. would like to thank the Frankfurt Institute for Advanced Studies (FIAS), Frankfurt am Main, Germany during the initial period of work on this project. P.N. and A.O. would like to thank the Perimeter Institute for Theoretical Physics, Waterloo, ON, Canada for the kind hospitality during the final period of work on this project.
\end{acknowledgments}

    \end{document}